# Secrecy Rate Study in Two-Hop Relay Channel with Finite Constellations

Zhen Qu, Shengli Zhang, Mingjun Dai, Hui Wang
Department of Communication Engineering, College of Information Engineering
Shenzhen University, Shenzhen China

*Abstract*—Two-hop security communication with an eavesdropper in wireless environment is a hot research direction. The basic idea is that the destination, simultaneously with the source, sends a jamming signal to interfere the eavesdropper near to or co-located with the relay. Similar as physical layer network coding, the friendly jamming signal will prevent the eavesdropper from detecting the useful information originated from the source and will not affect the destination on detecting the source information with the presence of the known jamming signal. However, existing investigations are confined to Gaussian distributed signals, which are seldom used in real systems. When finite constellation signals are applied, the behavior of the secrecy rate becomes very different. For example, the secrecy rate depends on phase difference between the input signals with finite constellations, which is not observed with Gaussian signals. In this paper, we investigate the secrecy capacity and derive its upper bound for the two-hop relay model, by assuming an eavesdropper near the relay and the widely used *M*-PSK modulation. With our upper bound, the best and worst phase differences in high SNR region are then given. Numerical studies verify our analysis and show that the derived upper bound is relatively tight.

*Keywords*—*M*-PSK; *finite constellations; capacity; secrecy rate; uppper bound;*

I. INTRODUCTION

Due to the broadcast nature of wireless communication, signal transmission in wireless environment is faced with a variety of security threats (e.g. interference, eavesdroppers, malicious node, untrusted relay, etc). Therefore, secrecy plays an important role in wireless networks and draws a lot of research attentions. Wyner investigated the wiretap channel [1] and showed that secure communication is possible in spite of the existence of an unauthorized intruder, under the condition that it can only observe degraded signals. For the general Gaussian multiple-access wiretap channel (GGMAC-WT) and the Gaussian two-way wiretap channel (GTW-WT), a corresponding secrecy measure is defined, and achievable secrecy rate regions are determined in [2]. A secure two-hop communication model with an untrusted relay is considered in [3-6]. In these works, the untrusted relay was confused by a friendly jamming signal to prevent the relay from decoding useful information originated from the source, where the jamming signal could be sent by the destination or an external friendly node. This idea is the same as the security mechanism of physical layer network coding [7-8].

We find that all the papers above were studied under the assumption of Gaussian distributed signal, including both the jamming and data signals. However, Gaussian distributed signals are rather theoretical and are seldom used in practical systems. More important, there is quite a little difference when practical finite constellations signals are adopted, especially taking the phase difference between the input signals into consideration. This scenario has only been noted in the area of Gaussian interference channel (GIC) and Gaussian multiple access channel. In [9], the two-user GIC with finite constellation-based transmitted signal is studied, and it is verified that if one user rotates its constellation appropriately with respect to the constellation of the other user, the achievable sum-rate in the network increases considerably. In [10], Constellation Constrained (CC) capacity region of two-user Single-Input Single-Output (SISO) Gaussian Multiple Access Channel are investigated under two schemes (non-orthogonal MA and orthogonal MA). To guarantee unique decodability property, appropriate rotation between the input constellations (such as *M*-PSK, *M*-QAM) is employed by the NO-MA scheme, with the objective of maximally enlarging the CC capacity. To the best of our knowledge, there is no work that discusses the finite constellation's effect on secure transmission.

In this paper, we adopt the two-hop relay channel model as shown in Fig. 1. We assume that there is an eavesdropper near the relay, which can overhear the transmissions from both the source and the relay. If we do not protect the transmitted signals, the eavesdropper has a high chance to recover the information when the signal strengths are high enough to guarantee correct detection at the destination. To keep the information secret from the eavesdropper[1], one effective way is to send an artificial interference from the destination to jamming the eavesdropper when the source is transmitting to the relay. This model is similar to the scenario where the relay node is untrusted [3-6], and the relay together with the eavesdropper co-located in the same entity.

Different from existing friendly interference works, which only consider Gaussian signal and jamming signal, we investigate secure transmissions where both the source and the destination only explore the widely used PSK modulations. This practical constellations assumption is not only reasonable in real wireless system, but also makes sense in theory. As a result, the secrecy rate is affected by the phase difference between the source and the destination, and this fact is not observed with Gaussian distributed signals.

---
[1] Note that the cryptograph schemes are not considered here since it has the probability to be deciphered in theory.

To derive the upper bound for secrecy rate, we apply the rate equivocation inequality in [11] and the information theory formulation. Due to PSK signals, when the phase difference is changed, the upper bound will be different. Since there is no closed-form expression for the upper bound of secrecy rate with varying phase difference, we only perform numerical study to reflect the change.

The rest of this paper is organized as follows: In section II, we present the system model; in section III, we derive the secrecy rate using information theory technique; in section IV, we derive an upper bound of the secrecy rate, and, we show the simulation results and conclusion in the last two sections, respectively.

## II. SYSTEM MODEL

Refer to Figure 1, S denotes the source, R denotes the relay, D denotes the destination, and E denotes the eavesdropper, which is adjacent to the relay. Solid lines represent the multiple-access (MA) phase, and dashed lines represent the broadcast (BC) phase. PSK-modulated signals $X_s$ and $X_d$ denote the signals from the source and the destination respectively, where $X_s$ contains confidential message, and $X_d$ is the jamming signal from the destination to the relay at the same time with $X_s$. We consider the scenario where the information originated from the source is not encrypted. Hence, we need a jamming signal to protect it from eavesdropping.

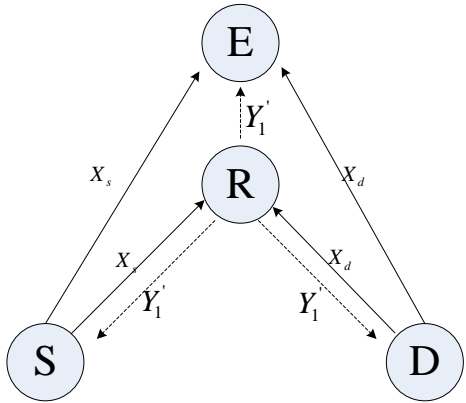

Fig.1. A two-hop wiretap channel

We assume that there are $n$ signals transmitted during the MA phase, and $m$ signals transmitted during the BC phase. We assume that the relay adopts amplify-and-forward (AF) scheme. We hence have $n = m$, and we use $W$ to represent the confidential message contained in $X_s$.

Because of PSK signals, we can let the constellation of signal adopt the $M$-PSK scheme. The source packet $X_s^n$ includes $n$ symbols $x_s^1, x_s^2 \cdots x_s^n$, with each symbol composed as $x_s^i = \sqrt{P_1} x_s$, where $P_1$ represents the power of the source, and $x_s$ is a $M$-PSK modulated signal that follows the uniform distribution in set $X \triangleq \left\{ e^{\frac{2\pi jm}{M}} : 0 \leq m \leq M-1 \right\}$. The interference packet $X_d^n$ is defined similarly:

$$x_d^i = e^{j\theta} \sqrt{P_2} x_d \quad (1)$$

where $P_2$ denotes the power of the destination and $\theta$ denotes the phase difference between the source and destination.

All the signals are transmitted in Gaussian channel. The length-$n$ vector signal received by the relay in the MA phase, denoted by $Y_1^n$, is formulated as:

$$Y_1^n = X_s^n + X_d^n + N_1^n \quad (2)$$

where the noise vector $N_1^n$ is a zero mean Gaussian random variable with unit variance.

At the same time, the signal that the eavesdropper received in the MA phase is denoted by $Y_e^n$ and is formulated as:

$$Y_e^n = X_s^n + X_d^n + N_2^n \quad (3)$$

where the noise vector $N_2^n$ is also a zero mean Gaussian random variable with unit variance. Hereafter, we ignore the superscript $n$, since the length of all vector signals is fixed.

In the BC phase, the relay broadcasts signal $Y_1'$, which is computed from $Y_1'$. We assume the relay adopts AF. So, we can get:

$$Y_1' = \alpha(X_s + X_d + N_1) \quad (4)$$

where the amplification factor $\alpha$ is a constant number either larger than 1 or smaller than 1, determined by the power constraint of the relay.

The signal received by the destination in the BC phase is denoted by $Y_2$, given by:

$$Y_2 = Y_1' + N_3 \quad (5)$$

where $N_3$ is a zero mean Gaussian random variable with unit variance.

Correspondingly, let $Y_e'$ denote the signal received by the eavesdropper during the BC phase, which is given by:

$$Y_e' = Y_1' + N_4 \qquad (6)$$

where $N_4$ is a zero mean Gaussian random variable with unit variance.

From (2), (3), and (5), we know that the eavesdropper receives two copies that carry the information. It can combine these two copies using the optimal maximal ratio combining (MRC). The final expression for the signal obtained by the eavesdropper is hence:

$$\begin{aligned}\overline{Y}_e &= X_s + X_d + \frac{\alpha(\alpha N_1 + N_4) + (\alpha+1)N_2}{\alpha^2 + \alpha + 1} \\ &= X_s + X_d + N_e\end{aligned} \qquad (7)$$

where $N_e = \dfrac{\alpha(\alpha N_1 + N_4) + (\alpha+1)N_2}{\alpha^2 + \alpha + 1}$ denotes a mixed Gaussian noise, with zero mean and variance $\sigma^2 = \dfrac{\alpha^4 + 2\alpha^2 + 2\alpha + 1}{(\alpha^2 + \alpha + 1)^2}$.

At the destination, the self-interference can be total removed from $Y_2$ and the final signal can be expressed by:

$$Y_d = X_s + N_1 + \frac{N_3}{\alpha} \qquad (8)$$

### III. SECRECY RATE

The eavesdropper will make full use of what it received to gain the confidential message. Hence the secrecy constraint can be denoted as:

$$\lim_{n+m\to\infty}\frac{1}{n+m}H(W) = \lim_{n+m\to\infty}\frac{1}{n+m}H(W|Y_e,Y_e') \qquad (9)$$

It means that the eavesdropper could obtain some information about $X_s$ containing the confidential message $W$, nonetheless, it can't get any useful information about $W$.

With satisfying of (9), the secrecy rate $R_e$ can be defined as:

$$R_e = \lim_{n+m\to\infty}\frac{1}{n+m}H(W) \qquad (10)$$

Due to the optimality of the MRC in (7), we can get:

$$R_e = \lim_{n+m\to\infty}\frac{1}{n+m}H(W) = \lim_{n+m\to\infty}\frac{1}{n+m}H(W|\overline{Y}_e) \qquad (11)$$

In Additive White Gaussian Noise (AWGN) wiretap channel, the secrecy capacity equals to the difference between the main and overall eavesdropper channel capacities [11]. From Shannon Theory, the achievable secrecy rate region is given as:

$$R_e \leq \max_{p(x)}[I(X_s;Y_d) - I(X_s;\overline{Y}_e)] \qquad (12)$$

where the right hand side of inequality (12) is wire-tap channel capacity according to [11], and it is maximized when $p(x)$ is Gaussian distribution[2]. In this system, for $M$-PSK modulation, each symbol is chosen with equal probability from the $M$ constellations and $p(x)$ is fixed. By spreading the right hand side of (12), we get:

$$\begin{aligned}R_e &\leq I(X_s;Y_d) - I(X_s;\overline{Y}_e) \\ &= h(X_s) - h(X_s|Y_d) - h(\overline{Y}_e) + h(\overline{Y}_e|X_s)\end{aligned} \qquad (13)$$

Since $\theta$ is the phase difference between $X_s$ and $X_d$, only $h(\overline{Y}_e)$ is correlated with $\theta$ in (13). To maximize the channel capacity, we should try our best to reduce $h(\overline{Y}_e)$ or $I(X_s;\overline{Y}_e)$, which can be achieved by adjusting the phase difference between the source and the destination.

### IV. UPPER BOUND

We assume the eavesdropper uses the Maximum-Likelihood (ML) detection method to get the transmitted symbol $X_s$ from $\overline{Y}_e$. In the following, we develop a lower bound on $I(X_s;\overline{Y}_e)$. Firstly, let

$$\hat{X}_s \triangleq \arg\max_{x\in X_s} p_{\overline{Y}_e|X_s}(\overline{Y}_e|x) \qquad (14)$$

We use $p_e = \Pr(\hat{X}_s \neq X_s)$ to denote the symbol error probability with the ML estimation in (14). Then, we have

$$\begin{aligned}I(X_s;\overline{Y}_e) &\overset{(a)}{\geq} I(X_s;\hat{Y}_e) = H(X_s) - H(X_s|\hat{Y}_e) \\ &\overset{(b)}{=} \log M - H(X_s|\hat{Y}_e) \\ &\overset{(c)}{\geq} \log M - H(p_e) - p_e\log(M-1)\end{aligned} \qquad (15)$$

---

[2] When the right hand side of (12) is negative, the secrecy rate is zero, and this extreme case is not considered in this paper.

where step (a) follows from the data processing inequality [12], and (b) is due to *M*-PSK constellations modulation, step (c) follows from Fano's inequality, and $H(p_e) = -p_e \log(p_e) - (1-p_e)\log(1-p_e)$ is a binary entropy. Therefore, we could obtain an upper bound as follows:

$$\begin{aligned}R_e &\leq I(X_s; Y_d) - I(X_s; \overline{Y_e}) \\ &\leq I(X_s; Y_d) - \log M + H(p_e) + p_e \log(M-1) \\ &= H(X_s) - H(X_s | Y_d) - \log M + H(p_e) + p_e \log(M-1) \quad (16)\\ &= \log M - H(X_s | Y_d) - \log M + H(p_e) + p_e \log(M-1) \\ &= H(p_e) + p_e \log(M-1) - H(X_s | Y_d)\end{aligned}$$

Note that $p_e$, which is the symbol error rate for the eavesdropper inferring the confidential information, depends on the phase difference $\theta$. The third term in (16), $H(X_s | Y_d)$, does not depends on $\theta$, and approaches to zero only if the destination decodes $X_s$ correctly. The first two terms of formula (16) are both correlated to $p_e$, and they serve a looser upper bound on the secrecy rate.

Our objective is to make $p_e$ as large as possible by changing $\theta$, so as to make the eavesdropper can't wiretap any useful information. Note that what he eavesdropper got is just the superposition of $X_s$, $X_d$, and Gaussian mixture noise $N_e$. Along with the angle changing, the virtual constellation of $X_s + X_d$ varies, so the minimum distance of the constellation, defined by $d_{\min}$, also changes, which will have effect on $p_e$. In high SNR region, it is widely known that the symbol error rate would become larger when the minimum distance becomes smaller, and vice versa. The highest and the lowest $p_e$ on $\theta$ can be respectively defined as:

$$\theta_h = \arg \min_\theta d_{\min} \quad (17)$$

and

$$\theta_l = \arg \max_\theta d_{\min} \quad (18)$$

By directly computing $d_{\min}(\theta)$, we could obtain $\theta_h$ and $\theta_l$ easily. For example, $\theta_h = 0$ and $\theta_l = \pi/6$ or $\pi/3$ when QPSK used at both source and destinations. At the same time, there is no doubt that it is more convenient for us to analyze $p_e$ and the upper bound. Then, we have the following conclusions:

***Proposition***: In high SNR region, the secrecy rate $R_e$ is maximized when $\theta = \theta_h$ and minimized when $\theta = \theta_l$.

A sketch of the proof is as follows. When SNR increase, the equality in Fano's inequality can be achieved and $H(p_e) + p_e \log(M-1)$ is an increasing function of $p_e$, which is a decreasing function of $d_{\min}$. The detailed proof is omitted due to limited space.

## V. SIMULATION RESULTS

In this section, we perform some numerical simulation results. We assume the signals of the source and the destination is QPSK modulated with the same power P1=P2=*P*.

Firstly, the minimum distance $d_{\min}$ and symbol error rate $p_e$ is shown in Fig.2. It is easy to see that the curve of symbol error rate $p_e$ is an inversion of minimum distance $d_{\min}$. $\theta_h$ is equal to $0$, and $\theta_l$ is equal to $\pi/6, \pi/3$, with a circle of $\pi/2$ for QPSK. With higher SNRs, this phenomenon is more obvious.

Fig.3 and Fig.4 shows the secrecy capacity and the upper bound with different phase differences and SNRs. With reference to Fig. 2, it is verified that the upper bound linearly increases with $p_e$. Furthermore, the upper bound becomes tighter when the power increases as shown in Fig.3 and Fig.4.

The behavior of the simulated secrecy rates are little different in Fig.3 and Fig. 4. When the SNR is 5 dB, the minimum rate is achieved at $\theta = \pi/4$ and the maximum rate is achieved at $\theta = 0$. Since in low SNR, the conclusion in our *Proposition* does not hold any more. In this case, it is more accurate to upper bound $R_e$ by regarding $X_s + X_d + N$ as a Gaussian distribution. When $\theta = \pi/4$, it is like Gaussian variable most and when $\theta = 0$, it is like Gaussian variable least.

When SNR increases to 10 dB, the curve of $R_e$ is almost the same as the upper bound and the conclusion in our *Proposition* holds.

As the SNR further increases as in Fig. 4, the distance between any two constellation points of $X_s + X_d$ is big enough for most phase differences. In other words, the eavesdropper can detect the information $(X_s, X_d)$ from $\overline{Y_e}$ easier and $h(\overline{Y_e})$ becomes larger. However, $h(\overline{Y_e})$ is limited by:

$$\begin{aligned}h(\overline{Y_e}) &= h(X_s + X_d + N_e) \\ &\leq H(X_s + X_d) + h(N_e) = 4 + h(N_e)\end{aligned} \quad (19)$$

which does not depend on the phase difference any more. This explains why the secrecy rate in Fig. 4 becomes flat.

Another interesting observation from Fig. 3 and Fig. 4 is that the power affects the secrecy rate in a non-monotonic way, which is different from the traditional point-to-point transmission. The secrecy rate is first increased when SNR changes from 5 dB to 10 dB as in Fig. 3 and it then decreased when SNR changes to 20 dB in Fig. 4. So, we need carefully design the transmission power to maximize the secrecy rate in our system.

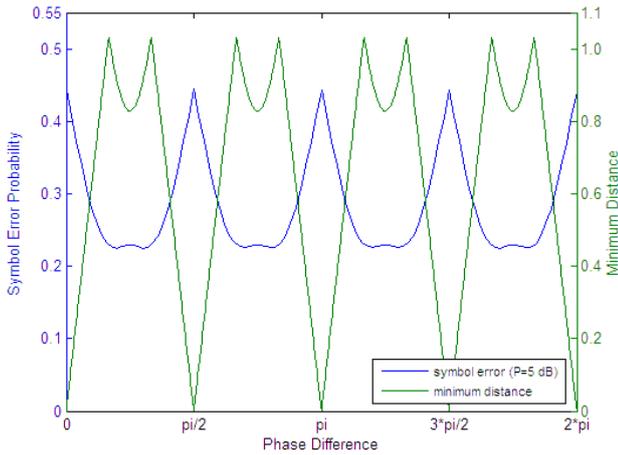

Fig.2. Plots of symbol error rate $p_e$ and minimum distance $d_{min}$ with the phase difference $\theta$.

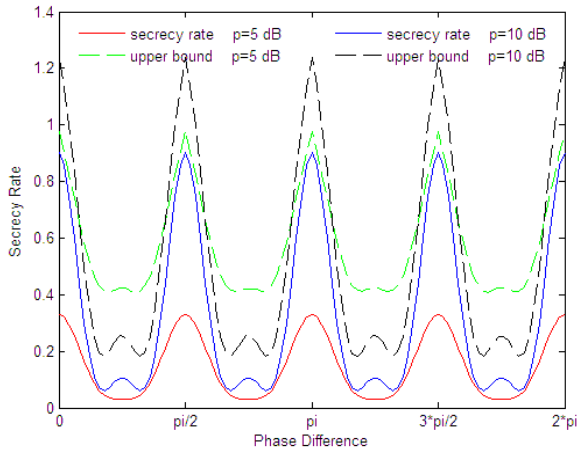

Fig.3. The secrecy rate and upper bound with the phase difference $\theta$.

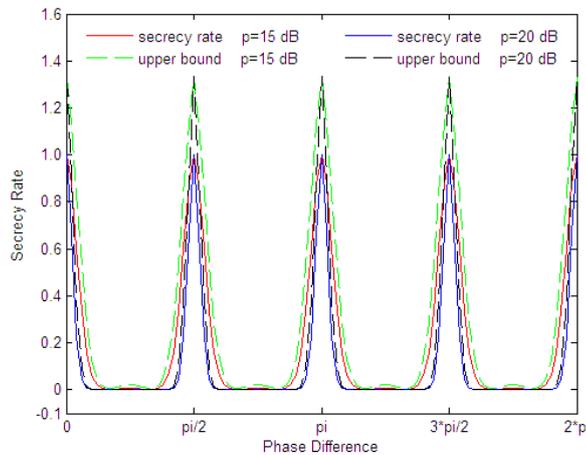

Fig.4. The secrecy rate and upper bound with phase difference $\theta$.

## VI. CONCLUSION

A two-hop secure Gaussian channel was studied based on finite constellations. It was found that when the different phase between the two users' (source node and destination node) constellations applied, the secrecy capacity varies. We gave the best and worst phase differences in an analytical way in high SNR region by deriving an upper bound. Numerical simulations not only verified our analysis but also showed the effect of SNR on secrecy rate. With the result in our paper, we can carefully design the transmission power and phase difference to maximize the system secrecy rate.